\documentclass[pre]{revtex4}

\usepackage{graphicx}
\usepackage{dcolumn}
\usepackage{bm}
\usepackage{subfigure}
\usepackage{amssymb}
\usepackage{graphics}
\usepackage{epsfig}
\usepackage{threeparttable}

\usepackage{threeparttable}

\begin{document}

\preprint{APS/123-QED}

\title{Nonequilibrium phase transitions and stationary state solutions of a three-dimensional random-field Ising model under a time dependent periodic external field }
\author{Yusuf Y\"{u}ksel}
\altaffiliation[Also at ]{Dokuz Eyl\"{u}l University,
Graduate School of Natural and Applied Sciences, Turkey}
\author{Erol Vatansever}
\altaffiliation[Also at ]{Dokuz Eyl\"{u}l University,
Graduate School of Natural and Applied Sciences, Turkey}
\author{\"{U}mit Ak{\i}nc{\i}}%
\affiliation{Department of Physics, Dokuz Eyl\"{u}l University,
TR-35160 Izmir, Turkey}
\author{Hamza Polat}
\email{hamza.polat@deu.edu.tr}
\affiliation{Department of Physics, Dokuz Eyl\"{u}l University,
TR-35160 Izmir, Turkey}


\date{\today}

\begin{abstract}
Nonequilibrium behavior and dynamic phase transition properties of a kinetic Ising model under the influence of periodically oscillating random-fields have been analyzed within the framework of effective field theory (EFT) based on a decoupling approximation (DA). Dynamic equation of motion has been solved for a simple cubic lattice ($q=6$) by utilizing a Glauber type stochastic process. Amplitude of the sinusoidally oscillating magnetic field is randomly distributed on the lattice sites according to bimodal and trimodal distribution functions. For a bimodal type of amplitude distribution, it is found in the high frequency regime that the dynamic phase diagrams of the system in temperature versus field amplitude plane resemble the corresponding phase diagrams of pure kinetic Ising model. Our numerical results indicate that for a bimodal distribution, both in the low and high frequency regimes, the dynamic phase diagrams always exhibit a coexistence region in which the stationary state (ferro or para) of the system is completely dependent on the initial conditions whereas for a trimodal distribution, coexistence region disappears depending on the values of system parameters.
\begin{description}
\item[\qquad \qquad \qquad PACS numbers]
03.65.Vf, 64.60.Ht, 75.10.Nr, 75.30.Kz
\end{description}
\end{abstract}

\maketitle
\tableofcontents

\section{Introduction}\label{intro}
Ising model in a quenched random magnetic field (RFIM) has attracted a considerable interest in the last three decades. The model which is actually based on the local fields acting on the lattice sites which are taken to be random according to a given probability distribution was introduced for the first time by Larkin \cite{larkin} for superconductors and later generalized by Imry and Ma \cite{imry_ma}. A lower critical dimension $d_{c}$ of the RFIM has remained an unsolved mystery for many years and now it is well established that a transition should exist in three and higher dimensions for
finite temperature and randomness, which means that $d_{c}=2$ \cite {imry_ma,grinstein,fernandez,imbrie,bricmont}. A great many experimental works have paid attention to the equilibrium properties of RFIM and quite noteworthy results have been obtained. For instance, it has been shown that diluted anti-ferromagnets such as $\mathrm{Fe_{x}Zn_{1-x}F_{2}}$ \cite{belanger,king}, $\mathrm{Rb_{2}Co_{x}Mg_{1-x}F_{4}}$ \cite{ferreira,yoshizawa}, and $\mathrm{Co_{x}Zn_{1-x}F_{2}}$ \cite{yoshizawa} in a uniform magnetic field just correspond to a ferromagnet in a random uniaxial magnetic field \cite{fishman,cardy}. From the theoretical point of view, equilibrium RFIM has been studied by a wide variety of techniques such as mean-field theory (MFT) \cite{schneider,andelman,aharony,mattis}, effective-field theory (EFT) \cite{borges, sarmento,sebastianes,kaneyoshi0,akinci}, Monte Carlo (MC) simulations \cite{landau,machta,fytas1,fytas2}, and the series expansion (SE) method \cite{gofman}. Based on these theoretical works, it is well known that different random-field distributions may lead to different phase diagrams and the presence of quenched randomness constitutes an important role in material science, since it may induce some important macroscopic effects on the thermal and magnetic properties of real materials.

When a ferromagnetic material is subject to a periodically varying time dependent magnetic field (kinetic Ising model), the system may not respond to the external magnetic field instantaneously which causes interesting behaviors due to the competing time scales of the relaxation behavior of the system and periodic external magnetic field. At high temperatures and for the high amplitudes of the periodic magnetic field, the system is able to follow the external field with some delay while this is not the case for low temperatures and small magnetic field amplitudes. This spontaneous symmetry breaking indicates the presence of a dynamic phase transition (DPT) \cite{chakrabarti} which shows itself in the dynamic order parameter (DOP) which is defined as the time average of the magnetization over a full period of the oscillating field. DPT properties of kinetic Ising model have been firstly observed theoretically by Tom\'{e} and Oliveira  within the framework of MFT \cite{tome}. Since then, much attention has been devoted to investigate the dynamic nature of the phase transitions by means of several theoretical and experimental works. On the theoretical side, non-equilibrium phase transition properties of kinetic Ising model  have been widely investigated by various techniques  \cite{pelcovitz,acharyya1,acharyya0,acharyya2,acharyya3,acharyya4,shi,deviren}. Besides, on the experimental picture, a DPT occurring for the high frequency magnetic fields was studied by Jiang et al. \cite{jiang} using the surface magneto-optical Kerr-effect technique for epitaxially grown ultrathin Co films on a Cu (001) surface. For a [Co($4\mathrm{A^{o}}$)/Pt($7\mathrm{A^{o}}$)] multi-layer system with strong perpendicular anisotropy, an example of DPT has been observed by Robb et al.  \cite{robb}. They found that the experimental non-equilibrium phase diagrams strongly resemble the dynamic behavior predicted from theoretical calculations of a kinetic Ising model. It is clear from these works, there exists a strong evidence of qualitative consistency between theoretical and experimental studies.

On the other hand, non-equilibrium stationary states and dynamic phase transition properties of RFIM have not been well understood yet, and there exists a limited number of works in the literature  \cite{hausmann,figueiredo,achharya5,crokidakis,costabile}. For instance, Hausmann et al. \cite{hausmann} considered the behavior of an Ising ferromagnet under the influence of a fast switching, random external field. According to the analytical results based on MFT for the stationary state of the system, they observed a novel type of first order phase transition which has also been verified by their extensive MC simulations. Paula et al. \cite{figueiredo} determined the stationary states of RFIM by using MFT and constructed the phase diagrams from the stationary states of the magnetization as a function of temperature and field amplitude. They found that the continuous phase transitions coincide with the equilibrium ones \cite{aharony}, while the first-order transitions occur at fields larger than the corresponding values at equilibrium. In addition, they also observed that the difference between the fields at the limit of stability of the ordered phase and that of the equilibrium  is maximum at zero temperature and vanishes at the tricritical point. Furthermore, Acharyya \cite{achharya5} studied the non-equilibrium dynamic phase transition, in the two-dimensional kinetic Ising model in the presence of a randomly varying magnetic field both by MFT and MC simulations and discerned that in contrast to the results found in Ref. \cite{hausmann}, the nature of the transition is always continuous. In a recent work, Crokidakis \cite{crokidakis} performed MC simulations on cubic lattices for a non-equilibrium Ising model that stochastically evolves under the simultaneous operation of several spin-flip mechanisms where the local magnetic fields change sign randomly with time due to competing kinetics. From the numerical results, it has been predicted that there exist first-order transitions at low temperatures and large disorder strengths, which correspond to the existence of a non-equilibrium tricritical point at finite temperature. Very recently, Costabile et al. \cite{costabile} have studied the dynamical phase transitions of the kinetic Ising model in the presence of a random magnetic field by using EFT with correlations where the EFT dynamic equation has been given for the simple cubic lattice $(q=6)$, and the dynamic order parameter has been calculated. It has been observed that the system presents ferromagnetic and paramagnetic states for low and high temperatures, respectively. Apart from this, they have predicted a non-equilibrium tricritical point in a phase diagram in the temperature versus applied field amplitude plane. They have also compared the results with the equilibrium phase diagram \cite{borges,sarmento}, where only the first-order line is different. In the theoretical works mentioned above, the random field effects have been taken into account either by a given probability distribution function (random in space) namely a bimodal distribution, or by generating a new configuration of random fields uniformly at each time step (random in time).

In the present paper, we have studied dynamic phase transitions and stationary states of RFIM driven by a periodically varying time dependent magnetic field on a simple cubic lattice. Amplitude of the applied magnetic field is sampled from both bimodal and trimodal probability distributions. For this purpose, we organized the paper as follows: In Sec. \ref{formulation} we briefly present the formulations. The results and discussions are summarized in Sec. \ref{results}, and finally Sec. \ref{conclusion} contains our conclusions.

\section{Formulation}\label{formulation}
We consider a three dimensional Ising ferromagnet $(J>0)$ defined on a simple cubic lattice with a time dependent external magnetic field. The time dependent Hamiltonian describing our model is
\begin{equation}\label{eq1}
\mathcal{H}=-J\sum_{<ij>}S_{i}S_{j}-\sum_{i}H_{i}(t)S_{i},
\end{equation}
where the first term is a summation over the nearest-neighbor spins with $S_{i}=\pm1$ and $H_{i}(t)$ is a time-dependent external oscillating magnetic field which is given by
\begin{equation}\label{eq2}
H_{i}(t)=H_{0i}\cos(\omega t),
\end{equation}
where $H_{0i}$ is the amplitude of the external magnetic field acting on the site $i$, and $\omega$ denotes the angular frequency of the oscillating external field. The amplitude of the field is distributed according to a given probability distribution function. The present study deals with a trimodal field distribution which has a form
\begin{equation}\label{eq3}
P(H_{0i})=p\delta(H_{0i})+\left(\frac{1-p}{2}\right)\left[\delta(H_{0i}-H_{0})+\delta(H_{0i}+H_{0})\right].
\end{equation}

In order to describe the dynamical evolution of the system, we follow a Glauber type stochastic process \cite{glauber}. The dynamical equation of motion can be obtained by using the master equation as follows
\begin{equation}\label{eq4}
\tau\frac{d\langle S_{i}\rangle}{dt}=-\langle S_{i}\rangle+\left\langle\tanh\left[\frac{E_{i}+H_{i}(t)}{k_{B}T}\right]\right\rangle,
\end{equation}
where $\tau$ is the transition rate per unit time, $E_{i}=J\sum_{j}S_{j}$ is the local field acting on the lattice site $i$, and $k_{B}$ and $T$ denote the Boltzmann constant and temperature, respectively.

In Eq. (\ref{eq4}), we set $\tau$ at unity. If we apply the differential operator technique \cite{honmura_kaneyoshi, kaneyoshi1} in Eq. (\ref{eq4}) by taking into account the random configurational averages we get
\begin{equation}\label{eq5}
\frac{dm}{dt}=-m+\left\langle\left\langle \prod_{j=1}^{q=6} \cosh(J\nabla)+S_{i}\sinh(J\nabla)\right\rangle\right\rangle_{r} F(x)|_{x=0},
\end{equation}
where $m=\langle\langle S_{i}\rangle\rangle_{r}$ represents the average magnetization, $\nabla=\partial/\partial x$ is a differential operator, $q$ is the coordination number of the lattice, and the inner $\langle...\rangle$ and the outer $\langle...\rangle_{r}$ brackets represent the thermal and configurational averages, respectively. The function $F(x)$ in Eq. (\ref{eq5}) is then defined by
\begin{equation}\label{eq6}
F(x)=\int dH_{0i}P(H_{0i})
\tanh\left[\frac{x+H_{i}(t)}{k_{B}T}\right].
\end{equation}
When the right-hand side of Eq. (\ref{eq5}) is expanded, the multispin correlation functions appear. The simplest approximation, and one of the most frequently adopted is to decouple these correlations according to
\begin{equation}\label{eq7}
\left\langle\left\langle
S_{i}S_{j}...S_{l}\right\rangle\right\rangle_{r}\cong\left\langle\left\langle
S_{i}\right\rangle\right\rangle_{r}\left\langle\left\langle
S_{j}\right\rangle\right\rangle_{r}...\left\langle\left\langle
S_{l}\right\rangle\right\rangle_{r},
\end{equation}
for $i\neq j \neq...\neq l$ \cite{tamura_kaneyoshi}.
If we expand the right-hand side of Eq. (\ref{eq5}) within the help of Eq. (\ref{eq7}) then we obtain the dynamical equation of motion as follows
\begin{equation}\label{eq8}
\frac{dm}{dt}=-m+\sum_{j=0}^{q=6}\Lambda_{j}m^{j}.
\end{equation}
The coefficients in Eq. (\ref{eq8}) are defined as
\begin{equation}\label{eq9}
\Lambda_{j}=\frac{1}{2^{q}} \sum_{r=0}^{q-j}\sum_{s=0}^{j}\left(\begin{array}{c}q-j \\ r \\ \end{array}\right)\left(\begin{array}{c} j \\ s \\ \end{array}\right)(-1)^{s}\exp\left[(q-2r-2s)J\nabla\right]F(x)|_{x=0},\quad j=0,1,...,q.
\end{equation}
These coefficients can be calculated by employing the mathematical relation $\exp(\alpha\nabla)F(x)=F(x+\alpha)$. Eq. (\ref{eq8}) can be regarded as a kind of initial value problem and the solution can be easily found benefiting from the initial value of the average order parameter $m_{0}$ by using the fourth order Runge-Kutta method (RK-4). For selected values of the Hamiltonian parameters and temperature, the time dependence of magnetization converges to a finite value after some iterations i.e. the solutions have property $m(t)=m(t+2\pi/\omega)$ for arbitrary initial value of the magnetization ($m_{0}$). Thus, by obtaining this convergent region after some transient steps (which depends on Hamiltonian parameters and the temperature) the DOP which is the time average of the magnetization over a full cycle of the oscillating magnetic field can be calculated from
\begin{equation}\label{eq10}
Q=\frac{\omega}{2\pi}\oint m(t)dt,
\end{equation}
where $m(t)$ is a stable and periodical function which can be one of the two types, according to whether it has the following
property or not \cite{tome}
\begin{equation}\label{eq11}
m(t)=-m(t+\pi/\omega).
\end{equation}
A solution satisfying Eq. (\ref{eq11}) is called symmetric solution which corresponds to a paramagnetic (P) phase where the magnetization oscillates around zero whereas the solution which does not ensure Eq. (\ref{eq11}) is called nonsymmetric solution, and it corresponds to ferromagnetic (F) phase where the magnetization oscillates around a non-zero value. In these two cases, the observed behavior of the magnetization is regardless of the choice of the initial value of magnetization. On the other hand, in contrast to the equilibrium RFIM, there exist coexistence regions (F+P phases) in the phase diagrams in temperature versus field amplitude plane where the stationary state of non-equilibrium RFIM problem depends on the initial value $m_{0}$ of the time dependent magnetization. Furthermore, it is not possible to obtain the free energy for kinetic models in the presence of time dependent external fields. Hence, in order to determine the type of the dynamic phase transition (first or second order), it is convenient to check the temperature dependence of DOP. Namely, if the DOP decreases continuously to zero in the vicinity of critical temperature, this transition is classified as of the second order whereas if it vanishes discontinuously then the transition is assumed to be of the first order.

\section{Results and discussion}\label{results}
In this section, we discuss how the random-fields effect the phase diagrams of the kinetic Ising model. Also, in order to clarify the type of the dynamic phase transitions in the system, we give the temperature dependence of the dynamic order parameter.

\begin{figure}[h!]
\center
\includegraphics[width=10cm]{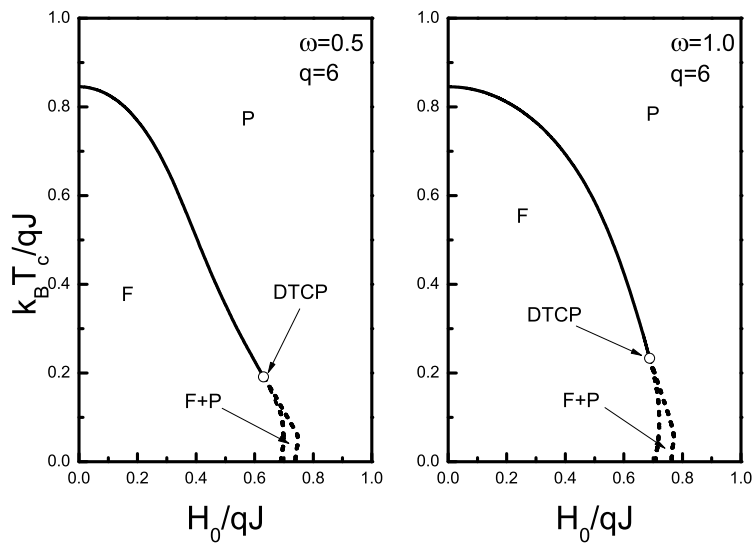}\\
\caption{Dynamic phase diagram of the pure kinetic Ising model in $(k_{B}T/qJ-H_{0}/qJ)$ plane for oscillation frequency values $\omega=0.5$ (left panel) and $\omega=1.0$ (right panel), for comparison with Refs. \cite{shi,deviren}. Solid (dashed) lines correspond to second (first) order phase transitions and solid symbols represent dynamic tricritcal points.}\label{fig1}
\end{figure}
\subsection{Non-equilibrium phase diagrams of the kinetic model: pure case}
In order to provide a testing ground for our calculations, we have primarily studied the dynamic phase diagrams of the kinetic Ising model under an oscillating magnetic field where the amplitude of the externally applied field was taken as a uniformly constant value. This model defines non-equilibrium properties of the pure system and has been examined previously within the framework of EFT \cite{shi,deviren}. In these works, the authors investigated the phase diagrams of the system in a $(k_{B}T_{c}/qJ-H_{0}/qJ)$ plane where $q$ is the coordination number of the lattice. From these works, we see that, for the oscillation frequency values $\omega=0.5$ and $1.0$, the location of dynamic tricritical (DTC) point was identified imprecisely. Hence, in order to compare our results with those found in Refs. \cite{shi,deviren}, we depict the phase diagrams in Fig. \ref{fig1} in the same plane. It is clear from Fig. \ref{fig1} that our numerical values of DTC point for $\omega=0.5$ and $1.0$ agree well with \cite{deviren} whereas they are quite different from those obtained in Ref. \cite{shi} which is probably due to the fact that in Ref. \cite{shi}, insufficient number of data points was used to construct the phase diagrams of the system.

\subsection{Non-equilibrium phase diagrams of RFIM for a bimodal distribution}
\begin{figure}[h!]
\center
\subfigure[\hspace{0 cm}] {\includegraphics[width=8.5cm]{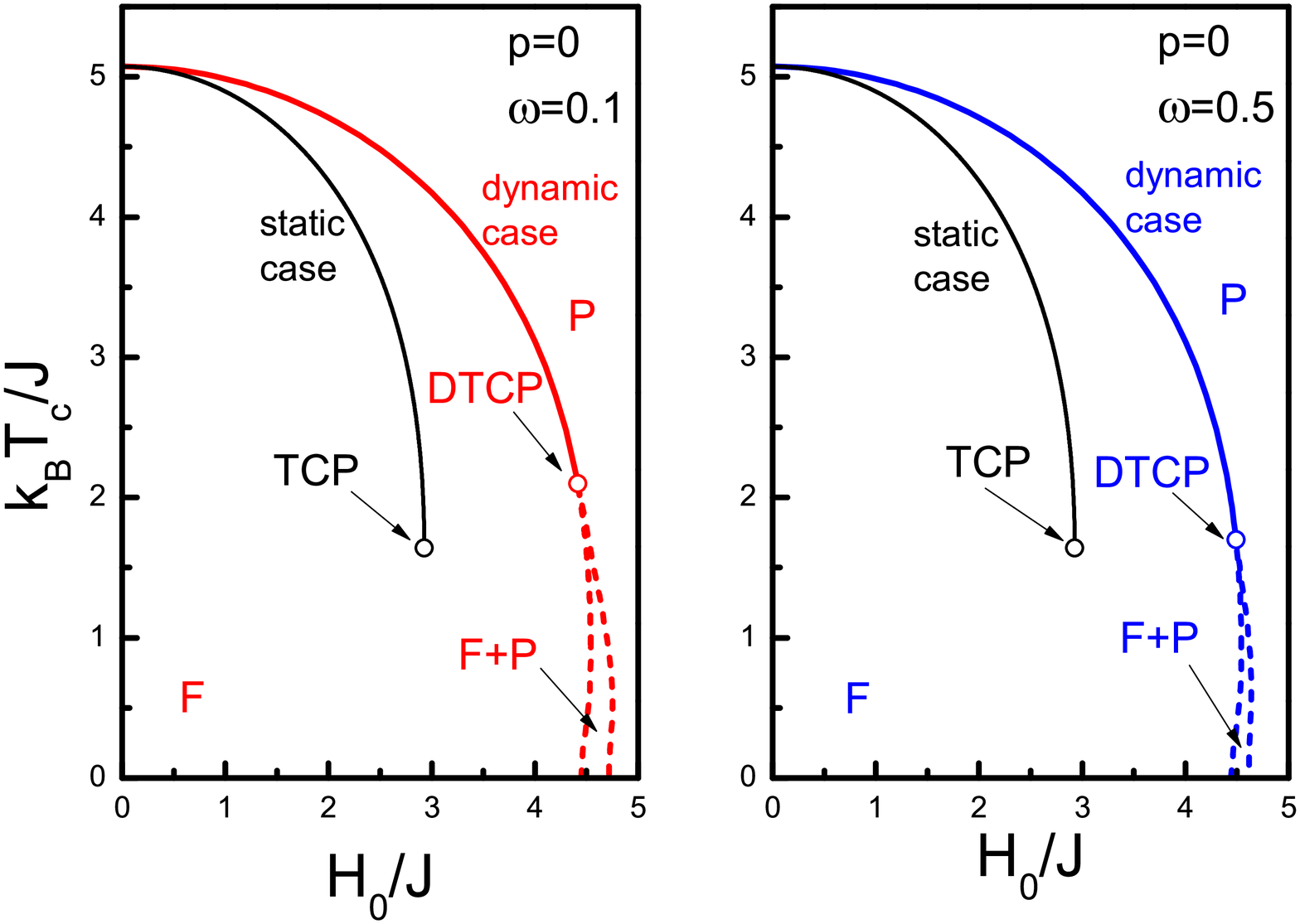}}
\subfigure[\hspace{0 cm}] {\includegraphics[width=8.5cm]{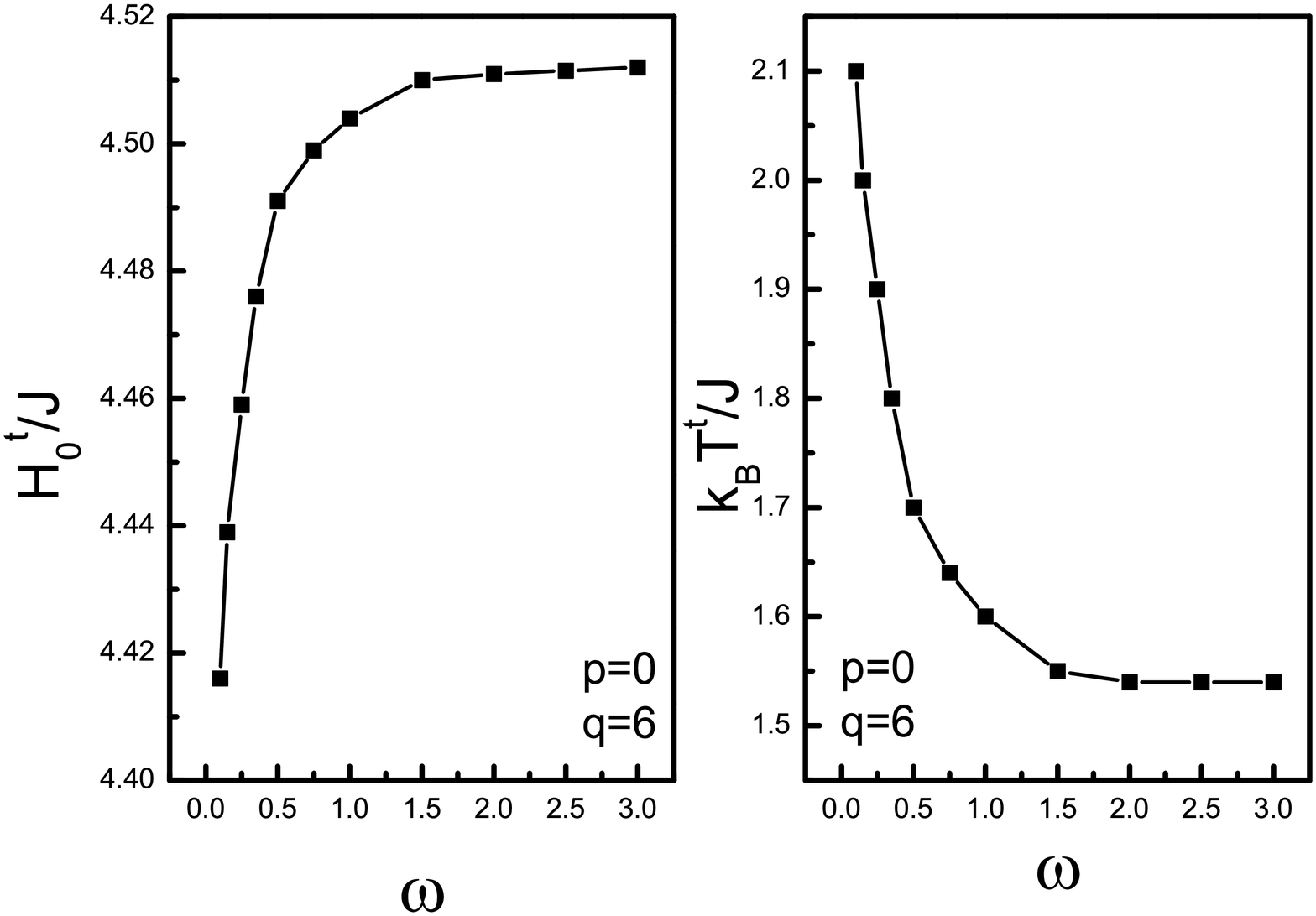}}\\
\caption{(Color online) (a) Phase diagrams of the kinetic RFIM in $(k_{B}T_{c}/J-H_{0}/J)$ plane for a bimodal distribution with $\omega=0.1$ and $0.5$, in comparison with the static model \cite{borges,sarmento}. Solid (dashed) lines correspond to second (first) order phase transitions and solid symbols represent tricritical points.     (b) Variation of the coordinates $H_{0}^{t}/J$ and $k_{B}T^{t}/J$ of dynamic tricritical point as a function of frequency $\omega$.}\label{fig2}
\end{figure}

The distribution function given in Eq. (\ref{eq3}) corresponds to a bimodal field distribution for $p=0$ where the amplitude of the oscillating field can be either $+H_{0}$ or $-H_{0}$ with equal probability. In this case, the system can be thought as a spin system under the influence of two oscillating external field sources. In Refs. \cite{figueiredo,costabile} a similar model has been studied within MFT and EFT, respectively where the authors did not consider any oscillating external magnetic field. Main conclusion of those studies was that the dynamic second order phase transition lines in $(k_{B}T_{c}/J-H_{0}/J)$ plane coincide with the equilibrium counterparts \cite{aharony,borges,sarmento} whereas the maximum and minimum differences between the dynamic and equilibrium first order phase transition lines were observed at the zero temperature and at the tricritical point, respectively. However, as seen in Fig. \ref{fig2}a, if the external magnetic field has an oscillatory character with an amplitude which is applied at random then a completely different situation arises. Namely, the second order phase transition fields of non-equilibrium RFIM with oscillating random external fields are greater than those obtained for static RFIM \cite{borges,sarmento}. On the other hand, according to the dynamic phase diagram shown in Fig. \ref{fig2}a, the system exhibits F phase for high temperature and low field amplitude values. In addition, as the oscillation frequency $\omega$ increases then DTC point depresses and the F+P phase region gets narrower. In Fig. \ref{fig2}b, variations of DTC point coordinates $H_{0}^{t}/J$ and $k_{B}T^{t}/J$ with respect to the frequency $w$ are plotted. As seen in this figure, $H_{0}^{t}/J$ grows whereas $k_{B}T^{t}/J$ decays with increasing $\omega$ and these values saturate at $H_{0}^{t}/J=4.512$ and $k_{B}T^{t}/J=1.54$. This means that the coexistence region in the phase diagrams gets narrower but does not disappear with increasing $\omega$. In other words, for sufficiently high frequencies, both the location of DTC point and the area of F+P region become independent of oscillation frequency of the external field. We also note that F region is always independent of frequency for the whole range of $\omega$.

\begin{figure}[h!]
\center
\includegraphics[width=10cm]{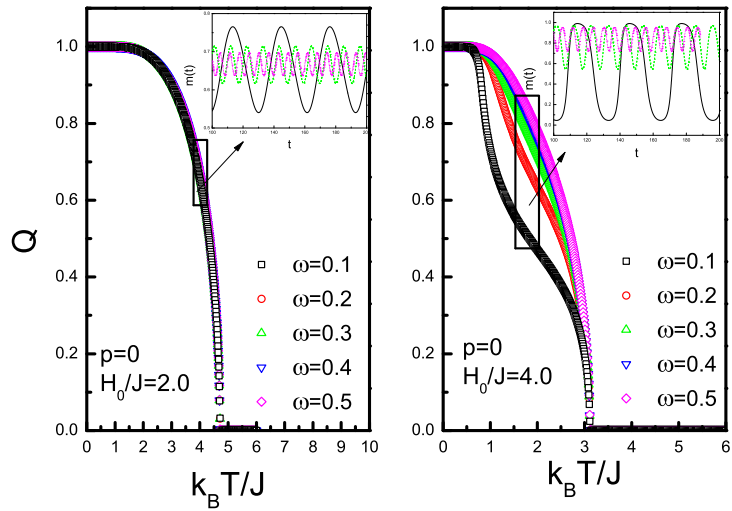}\\
\caption{(Color online) Temperature dependence of the dynamic order parameter $Q$ as a function of frequency $\omega$ for $H_{0}/J=2.0$ (left panel) and 4.0 (right panel). Insets in both panels show the time variation of the average magnetization $m(t)$ at $k_{B}T/J=4.0$ (left) and $1.75$ (right).}\label{fig3}
\end{figure}

In Fig. \ref{fig3}, we show the temperature dependence of the dynamic order parameter $Q$ as a function of the oscillation frequency $\omega$ corresponding to the phase diagrams depicted in Fig. \ref{fig2}. As we mentioned before, increasing or decreasing frequency values does not affect the area of the F region. Namely, for a fixed $H_{0}/J$ value, the system always undergoes a dynamic phase transition on the F-P phase boundary line, and the critical temperature value is independent of $\omega$. This situation is independent of the selection of the initial magnetization for $H_{0}/J<H_{0}^{t}/J$ whereas it is valid for $H_{0}/J>H_{0}^{t}/J$ only if one starts with an initial condition $m_{0}=0$. As is clearly seen in Fig. \ref{fig3}, all curves coincide with each other for $H_{0}/J=2.0$. The inset shows the average magnetization $m(t)$ as a function of time $t$ at $k_{B}T/J=4.0$ and for $\omega=0.1, 0.3$ and $0.5$ where we see that $m(t)$ curve oscillates with smaller amplitude as $\omega$ increases, but the average value, i.e. DOP $(Q)$ over a complete cycle of the magnetic field does not change. On the other hand, as shown in the right panel of Fig. \ref{fig3}, for sufficiently strong amplitudes of the external field, such as $H_{0}/J=4.0$, DOP versus temperature curves exhibit different characteristics  with increasing frequency, although the critical temperature does not change, and interestingly, for sufficiently high frequency values such as $\omega\geq0.5$ DOP curves resemble those of the pure kinetic Ising model driven by a periodic external field with the high oscillation frequency.

\begin{figure}[h!]
\center
\includegraphics[width=14cm]{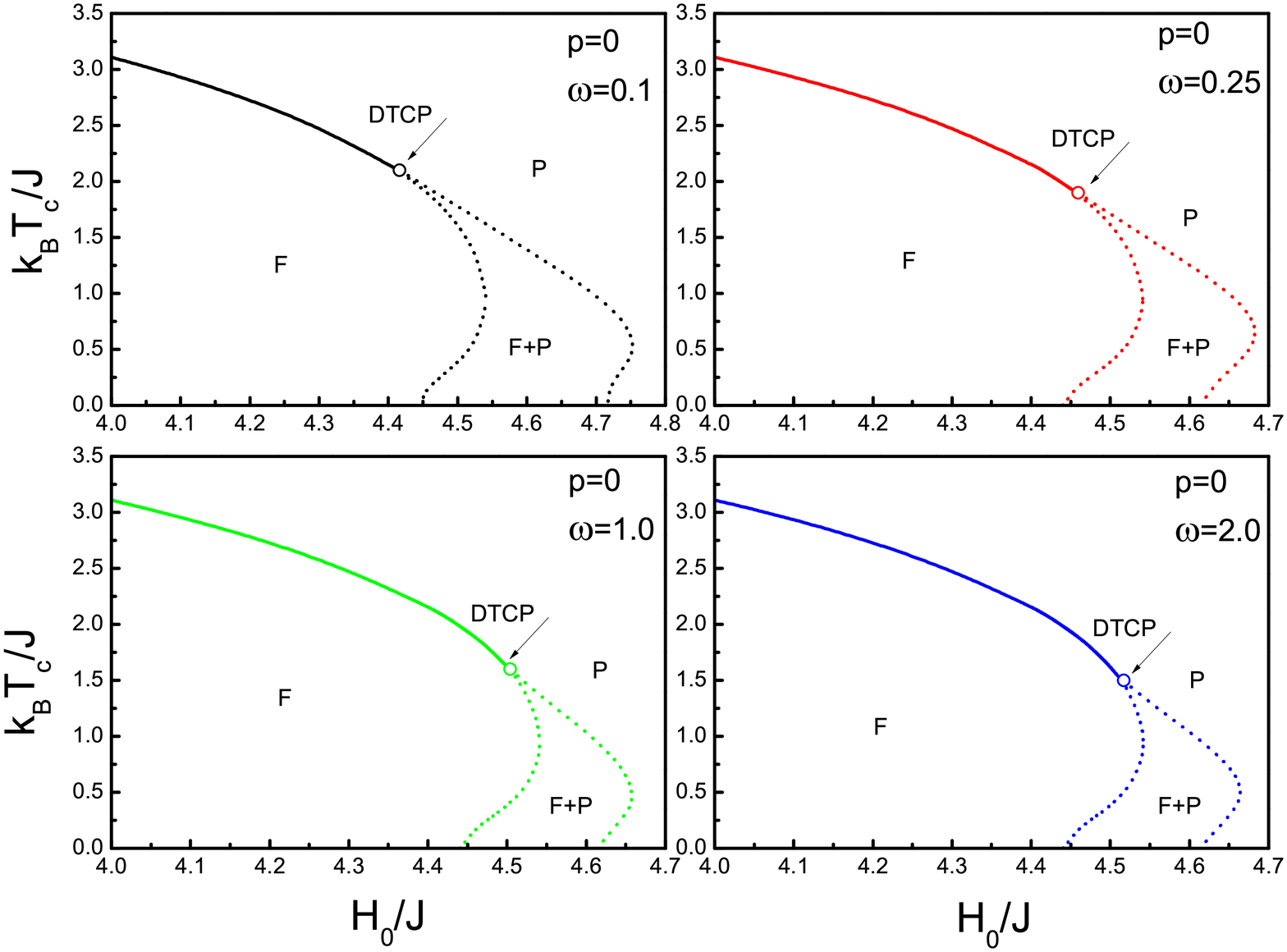}\\
\caption{(Color online) Dynamic phase diagrams of the system in a $(k_{B}T_{c}/J-H_{0}/J)$ plane for a bimodal distribution with some selected values of frequency $\omega$. The solid (dashed) lines correspond to second (first) order phase transitions and solid symbols represent tricritical points.}\label{fig4}
\end{figure}
In Fig. \ref{fig4}, dynamic phase diagrams of the system are depicted for a wide range of oscillation frequency values. In this figure, we observe that DTC point depresses for a while and the area of the coexistence region in the phase diagrams slightly gets narrower and then remains unchanged with increasing frequency values. The low frequency phase diagrams of the system are found to be completely different from those of the pure kinetic Ising model \cite{deviren} whereas for the high frequency regime, dynamic phase diagrams of the system are exactly identical to those of the pure case. These observations originate from the symmetry of the random field distribution. Namely, for a bimodal field distribution, we have two oscillating magnetic field sources. Initially, half of the lattice sites are under the influence of a periodically oscillating field (source-1) with an amplitude $H_{0}/J$, and the remaining spins on the other half of the lattice are influenced by an oscillating external field with an amplitude $-H_{0}/J$ (source-2). As the time progresses, due to the sinusoidal form of the periodic fields, magnetic field sources acting on the lattice sites change their sign instantaneously at the end of the one half of the oscillation period. Consequently, the presence of a phase difference between the source-1 and source-2 stimulates some unusual effects on the system. Furthermore, for sufficiently high frequencies of the oscillating external fields, relaxation time of the system becomes greater than the oscillation period of the external field. As a result of this, source-1 and source-2 fields receive identical responses from the system which perceive the field sources as a single oscillatory field source, hence the dynamic phase diagrams of the system resemble the high frequency phase diagrams of the corresponding pure model. We may also note that F+P regions in the $k_{B}T_{c}/J-H_{0}/J$ phase diagrams always exist for the whole range of frequency values.

\begin{figure}[h!]
\center
\includegraphics[width=10cm]{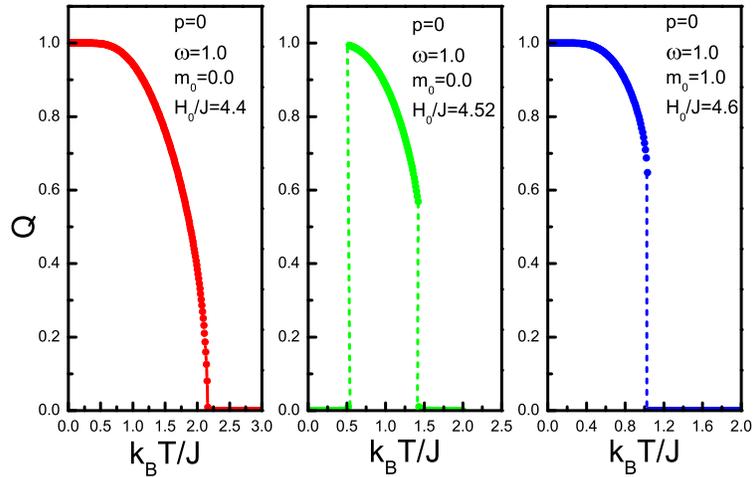}\\
\caption{(Color online) Variation of the dynamic order parameter $Q$ as a function of the temperature for $\omega=1.0$ with some selected values of $H_{0}/J$. The dashed lines correspond to first order phase transitions.}\label{fig5}
\end{figure}

In Fig. \ref{fig5}, we represent some examples of typical DOP versus temperature profiles corresponding to the phase diagrams shown in Fig. \ref{fig4} with $\omega=1.0$. Namely, the system always undergoes a second-order dynamic phase transition for $H_{0}/J=4.4$. In this case, dynamic nature of the phase transition and the stationary state of the system are independent of the initial magnetization $m_{0}$. On the other hand, two successive first order dynamic phase transitions (i.e., a first-order reentrant phenomena) occur at the value $H_{0}/J=4.52$ with the initial condition $m_{0}=0.0$. In addition, for $H_{0}/J=4.6$ with $m_{0}=1.0$, DOP curve exhibit a discontinuous jump at a phase transition temperature which puts forward the existence of a first-order transition on the system. For the latter two examples, dynamic nature of the phase transition and the stationary state of the system strictly depends on the initial magnetization.

\subsection{Non-equilibrium phase diagrams of RFIM for a trimodal distribution}
\begin{figure}[h!]
\subfigure[\hspace{0 cm}] {\includegraphics[width=8cm]{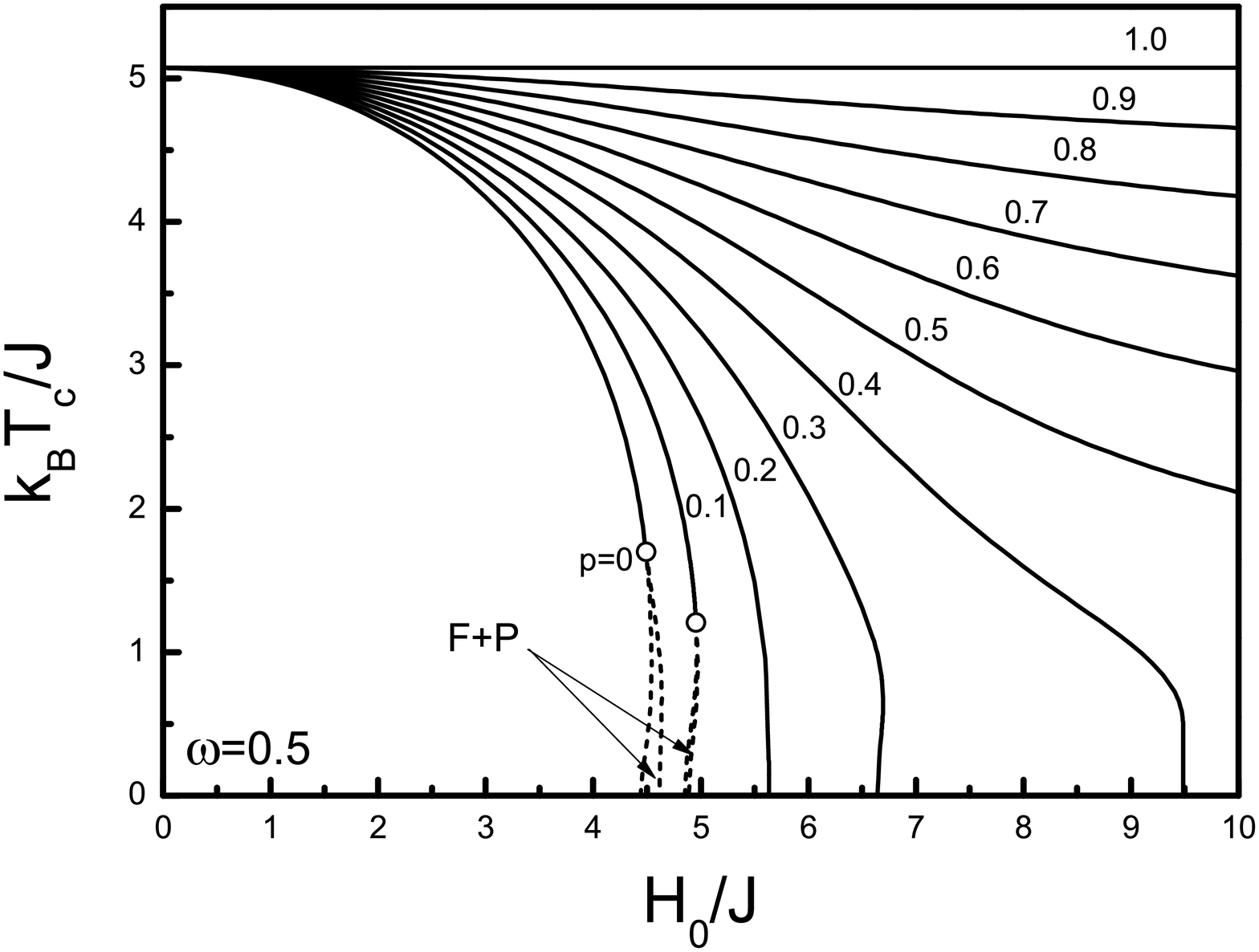}}
\subfigure[\hspace{0 cm}] {\includegraphics[width=8cm]{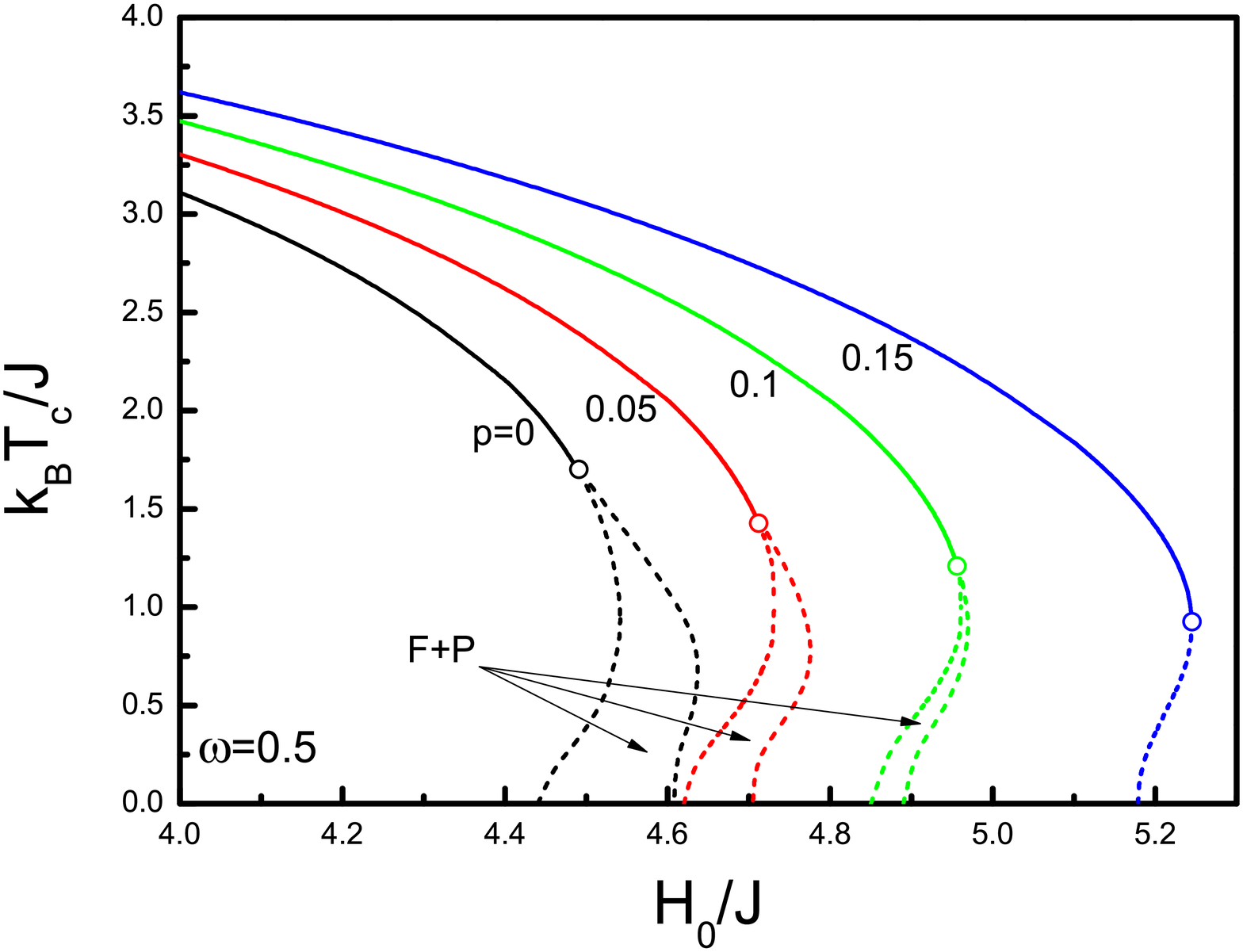}}\\
\caption{(Color online) Dynamic phase diagrams of the system in a $(k_{B}T_{c}/J-H_{0}/J)$ plane for a trimodal distribution of random fields with $\omega=0.5$ and for some selected values of $p$. The solid (dashed) lines correspond to second (first) order phase transitions and symbols represent tricritical points.}\label{fig6}
\end{figure}

For a trimodal field distribution of magnetic fields defined in Eq. (\ref{eq3}), we plot the dynamic phase diagrams of the system in a $(k_{B}T_{c}/J-H_{0}/J)$ plane in Fig. \ref{fig6} for $\omega=0.5$ and with some selected values of disorder parameter $p$. As shown in Fig. \ref{fig6}a, the system exhibits DTC behavior for a relatively narrow range of $p$. The temperature coordinate $k_{B}T^{t}/J$ of DTC point decreases as $p$ increases whereas the field amplitude part $H_{0}^{t}/J$ increases, and consequently DTC point depresses to zero at a certain value of $p$. After the destruction of the DTC point, all dynamic phase transition processes are found to be of the second order. We also note that at the zero temperature, there exist a critical value $p^{*}=0.53$ below which the system exhibits a dynamic phase transition at a critical magnetic field. This value of $p^{*}$ is independent of $\omega$. Moreover, according to Fig. \ref{fig6}b, as $p$ increases then the area of F+P region in the dynamic phase diagrams gets narrower and after a specific value of $p$, we cannot observe any coexistence region in $(k_{B}T_{c}/J-H_{0}/J)$ plane.

\begin{figure}[h!]
\center
\includegraphics[width=10cm]{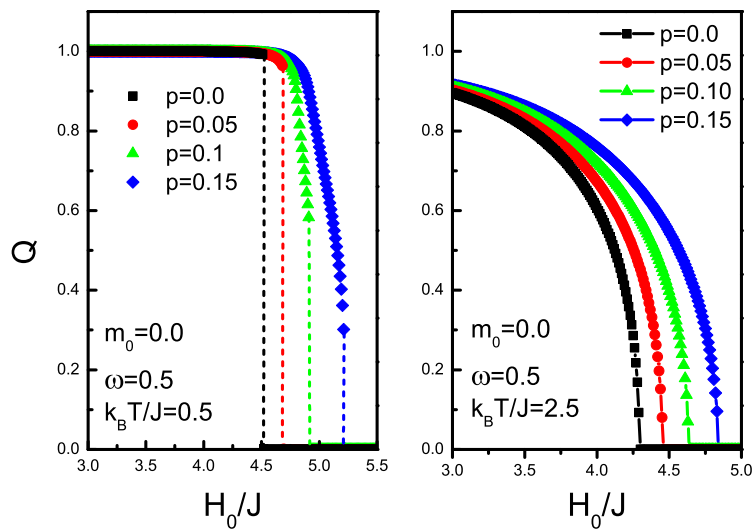}\\
\caption{(Color online) Variation of DOP $(Q)$ as a function of the amplitude of the oscillating field $H_{0}/J$ for a trimodal distribution corresponding to the dynamic phase diagrams depicted in Fig. \ref{fig6} with some selected values of $p$ for $m_{0}=0$, $\omega=0.5$ and for the reduced temperatures (a) $k_{B}T/J=0.5$, (b) $k_{B}T/J=2.5$. }\label{fig7}
\end{figure}
Variation of DOP $(Q)$ as a function of the amplitude of the oscillating field $H_{0}/J$ for a trimodal distribution corresponding to the dynamic phase diagrams depicted in Fig. \ref{fig6} is plotted in Fig. \ref{fig7} with some selected values of $p$. In this figure, initial value of the average magnetization is selected as $m_{0}=0.0$. According to Fig. \ref{fig7}, DOP curves undergo first and second order dynamic phase transitions at low and high temperatures, respectively. At low temperatures, critical amplitude value $H_{0}^{c}/J$ at which a dynamic phase transition occurs depends on the initial magnetization $m_{0}$. Namely, for $m_{0}=0.0$ critical $H_{0}^{c}/J$ value is greater than that obtained for $m_{0}=1.0$ which is due to the presence of F+P phase in the system at a fixed value of $p$ (see Fig. \ref{fig6}b).

\begin{figure}[h!]
\center
\subfigure[\hspace{0 cm}] {\includegraphics[width=8cm]{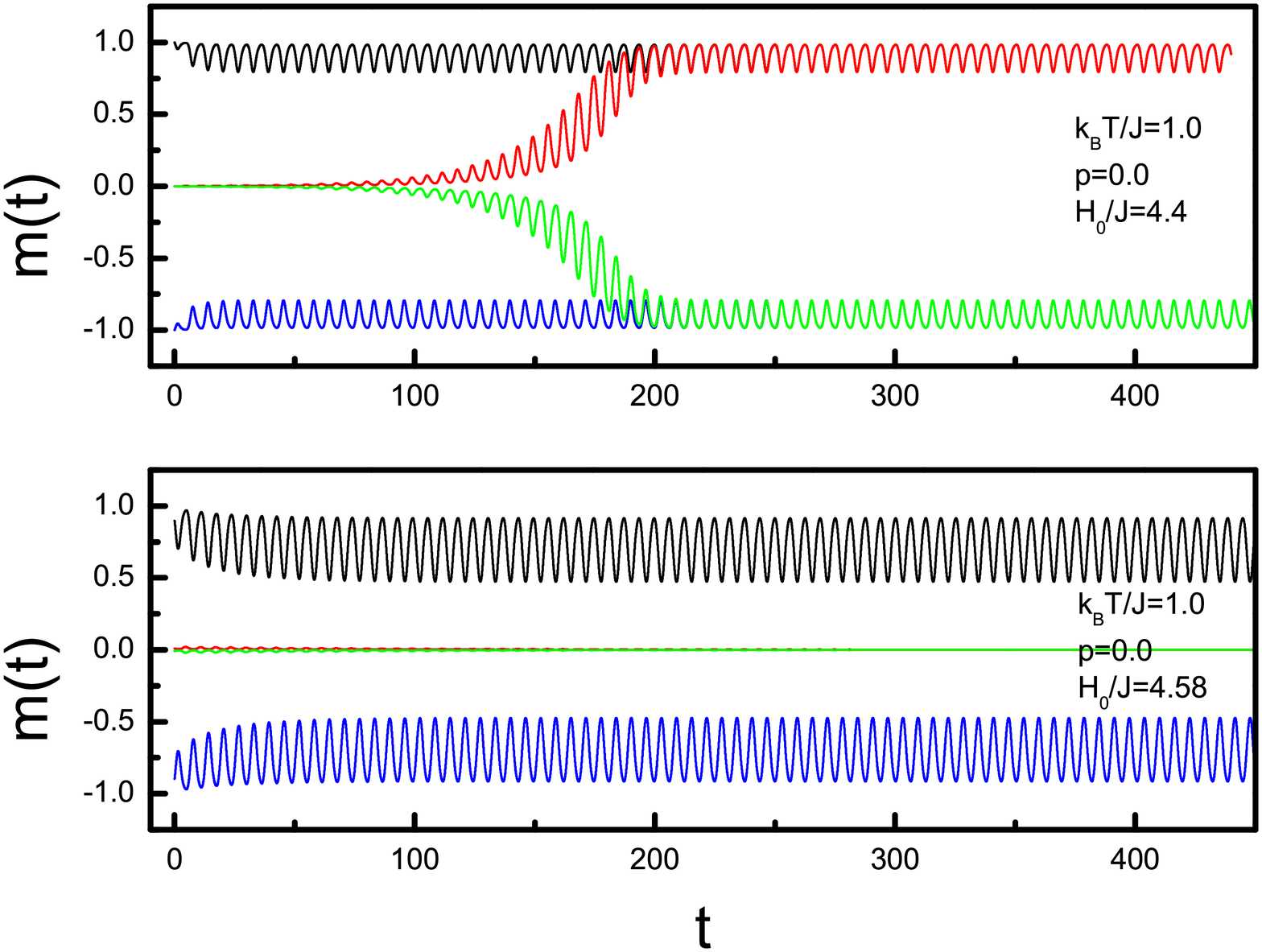}}
\subfigure[\hspace{0 cm}] {\includegraphics[width=8.3cm]{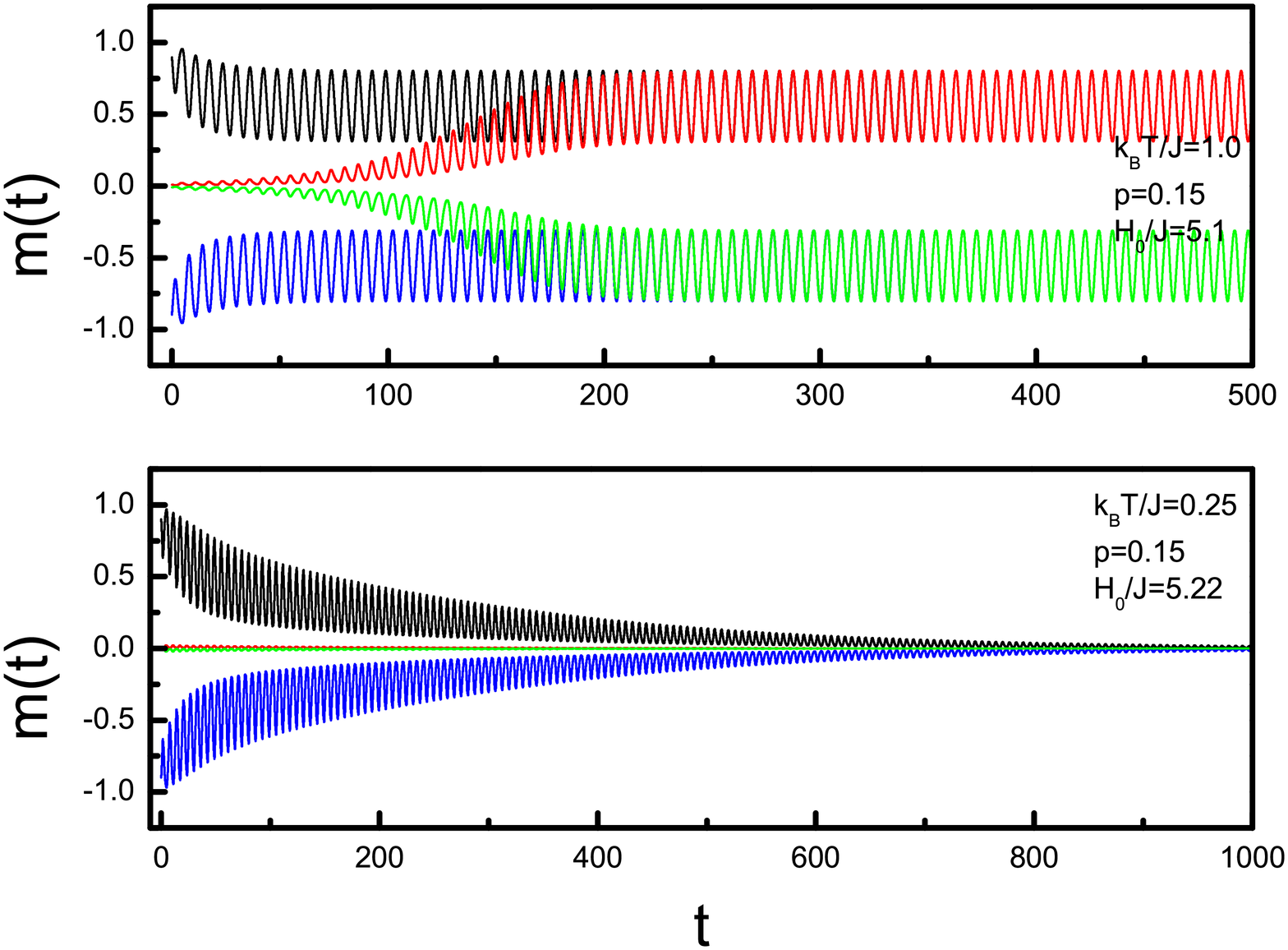}}\\
\caption{(Color online) Time dependence of the average magnetization $m(t)$ for $\omega=0.5$ with some selected values of the parameters $p$, $k_{B}T/J$, and $H_{0}/J$.}\label{fig8}
\end{figure}

Finally, for a fixed oscillation frequency $\omega=0.5$, let us investigate the time dependence of the average magnetization $m(t)$ corresponding to the phase diagrams shown in Fig. \ref{fig6}b. In Fig. \ref{fig8}a, we plot the time series of $m(t)$ curves for $p=0$, $k_{B}T/J=1.0$, and with two values of field amplitude $H_{0}/J$. From the upper panel in Fig. \ref{fig8}a, we see that the dynamic nature of the stationary state of the system is independent of the initial magnetization $m_{0}$ for $H_{0}/J=4.4$. Namely, for this set of system parameters, average magnetization of the system oscillates around a nonzero value (F phase) after some transient time. However, for $H_{0}/J=4.58$ (while the other parameters are the same) stationary state of the system (F or P) depends on the initial magnetization $m_{0}$ which indicates that the dynamic phase diagrams exhibit a coexistence region (F+P phase) in $(k_{B}T_{c}/J-H_{0}/J)$ plane. On the other hand, as seen in Fig. \ref{fig8}b which is plotted for $p=0.15$, we cannot observe any coexistent phase region in the system. Namely, for $H_{0}/J=5.1$ with $k_{B}T/J=1.0$, $m(t)$ curves oscillate around a finite nonzero value, whereas for $H_{0}/J=5.22$ and $k_{B}T/J=0.25$ the curves oscillates around zero which means that the system is in P phase. Hence, we see that the stationary state of the system for a trimodal distribution of the oscillating field amplitude may be independent of $m_{0}$, depending on the value of distribution parameter $p$.

\section{Concluding remarks}\label{conclusion}
In conclusion, we have investigated the kinetic behavior of a spin-1/2 Ising model on a simple cubic lattice $(q=6)$ under the influence of a quenched random magnetic field which oscillates periodically in time by means of an effective-field theory based on a standard decoupling approximation and the time evolution of the system has been presented by utilizing a Glauber type stochastic process. For bimodal and trimodal field distributions, we have studied the global dynamic phase diagrams of the system in a $(k_{B}T_{c}/J-H_{0}/J)$ plane and variation of the dynamic order parameter with respect to the temperature and random amplitude of the oscillating field, as well as the time series of the average magnetization curves. We found that the dynamic behavior of the random field Ising model with a periodically oscillating magnetic field exhibit quite different characteristics in comparison with the static RFIM problem.

Our numerical analysis clearly indicates that such field distributions may lead to a number of interesting phenomena.  For example, we have found that for a bimodal distribution with sufficiently high frequency values, DOP curves and non-equilibrium phase diagrams of the system resemble those of the pure kinetic Ising model driven by a periodic external field with the high oscillation frequency. On the other hand, we have also observed that F+P regions in $(k_{B}T_{c}/J-H_{0}/J)$ phase diagrams always exist for the whole range of frequency values for a bimodal distribution of oscillatory field amplitude whereas for a trimodal distribution, area of F+P region gets narrower with increasing distribution parameter $p$,  and if we increase $p$ further then we cannot observe any coexistence region in $(k_{B}T_{c}/J-H_{0}/J)$ plane.

EFT method takes the standard mean field predictions one step forward by taking into account the single site correlations which means that the thermal
fluctuations are partially considered within the framework of EFT. Although all of the observations reported in this work shows that EFT can be successfully applied to such nonequilibrium systems in the presence of quenched disorder, the true nature of the physical facts underlying the observations displayed in the system (especially the origin of the coexistence phase) may be further understood with an improved version of the present EFT formalism which can be achieved by attempting to consider the multi site correlations which originate when expanding the spin identities.
We believe that this attempt could provide a treatment beyond the present approximation.

\section*{Acknowledgements}
One of the authors (Y.Y.) would like to thank the Scientific and Technological Research Council of Turkey (T\"{U}B\.{I}TAK) for partial financial support. The numerical calculations reported in this paper were performed at T\"{U}B\.{I}TAK ULAKBIM, High Performance and Grid Computing Center (TR-Grid e-Infrastructure).


\end{document}